

\input boardmacs

\def\bra#1{{\langle#1\vert}}
\def\ket#1{{\vert#1\rangle}}
\def\coeff#1#2{{\scriptstyle{#1\over #2}}}

\def\hcal#1{{\hbox{\cal #1}}}
\def\sst#1{{\scriptscriptstyle #1}}

\def\rbra#1{{\langle #1 \vert\!\vert}}
\def\rket#1{{\vert\!\vert #1\rangle}}

\def\alr{{A_\sst{LR}}}

\def\evec{{\vec e}}
\def\notq{{\not\! q}}

\def\mns{{m^2_\sst{N}}}
\def\me{{m_e}}
\def\mes{{m^2_e}}

\def\mt{{m_t}}

\def\mzs{{M^2_\sst{Z}}}

\def\mh{{M_\sst{H}}}

\def\sbar{{\bar s}}

\def\sstw{{\sin^2\theta_\sst{W}}}

\def\qw{{Q_\sst{W}^2}}

\def\gve{{g_\sst{V}^e}}
\def\gae{{g_\sst{A}^e}}

\def\qv{{\vec q}}

\def\xv{{\vec x}}

\def\sqr#1#2{{\vcenter{\vbox{\hrule height.#2pt
                \hbox{\vrule width.#2pt height#1pt \kern#1pt
                        \vrule width.#2pt}
                \hrule height.#2pt}}}}

\def\hcal#1{{\hbox{\cal #1}}}
\def\sst#1{{\scriptscriptstyle #1}}

\def\notq{{\not\! q}}

\def\mns{{m^2_\sst{N}}}

\def\Hhat{{\hat H}}

\def\sst#1{{\scriptscriptstyle #1}}
\def\hcal#1{{\hbox{\cal #1}}}

\def\Hhat{{\hat H}}

\def\mns{{m_\sst{N}^2}}

\def\qv{{\vec q}}

\def\xv{{\vec x}}

\def\notq{{\rlap/q}}

\def\rbra#1{{\langle#1\parallel}}
\def\rket#1{{\parallel#1\rangle}}

\def\xivz{{\xi_\sst{V}^{(0)}}}

\def\rvz{{R_\sst{V}^{(0)}}}

\def\GES{{G_\sst{E}^{(s)}}}
\def\GMS{{G_\sst{M}^{(s)}}}

\def\mustr{{\mu_s}}

\def\rhostr{{\rho_s}}

\def\GEn{{G_\sst{E}^n}}
\def\GEp{{G_\sst{E}^p}}
\def\GMn{{G_\sst{M}^n}}

\def\lamd{{\lambda_\sst{D}^\sst{V}}}
\def\lamn{{\lambda_n}}
\def\lams{{\lambda_\sst{E}^{(s)}}}

\def\Gdip{{G_\sst{D}^\sst{V}}}

\def\RVp{{R_\sst{V}^p}}
\def\RVn{{R_\sst{V}^n}}

\def\bra#1{{\langle#1\vert}}
\def\ket#1{{\vert#1\rangle}}
\def\coeff#1#2{{\scriptstyle{#1\over #2}}}

\def\hcal#1{{\hbox{\cal #1}}}
\def\sst#1{{\scriptscriptstyle #1}}

\def\rbra#1{{\langle #1 \vert\!\vert}}
\def\rket#1{{\vert\!\vert #1\rangle}}

\def\alr{{A_\sst{LR}}}

\def\evec{{\vec e}}
\def\notq{{\not\! q}}

\def\mns{{m^2_\sst{N}}}
\def\me{{m_e}}
\def\mes{{m^2_e}}

\def\mt{{m_t}}

\def\mzs{{M^2_\sst{Z}}}

\def\mh{{M_\sst{H}}}

\def\sbar{{\bar s}}

\def\sstw{{\sin^2\theta_\sst{W}}}

\def\qw{{Q_\sst{W}^2}}

\def\gve{{g_\sst{V}^e}}
\def\gae{{g_\sst{A}^e}}

\def\qv{{\vec q}}

\def\xv{{\vec x}}

\def\sqr#1#2{{\vcenter{\vbox{\hrule height.#2pt
                \hbox{\vrule width.#2pt height#1pt \kern#1pt
                        \vrule width.#2pt}
                \hrule height.#2pt}}}}

\def\hcal#1{{\hbox{\cal #1}}}
\def\sst#1{{\scriptscriptstyle #1}}

\def\notq{{\not\! q}}

\def\mns{{m^2_\sst{N}}}

\def\Hhat{{\hat H}}

\def\sst#1{{\scriptscriptstyle #1}}
\def\hcal#1{{\hbox{\cal #1}}}

\def\Hhat{{\hat H}}

\def\mns{{m_\sst{N}^2}}

\def\qv{{\vec q}}

\def\xv{{\vec x}}

\def\notK{{\rlap/K}}
\def\notKp{{{\notK}^{\>\prime}}}
\def\nots{{\rlap/s}}

\def\notq{{\rlap/q}}

\def\rbra#1{{\langle#1\parallel}}
\def\rket#1{{\parallel#1\rangle}}

\def\xivz{{\xi_\sst{V}^{(0)}}}

\def\rvz{{R_\sst{V}^{(0)}}}

\def\GES{{G_\sst{E}^{(s)}}}
\def\GMS{{G_\sst{M}^{(s)}}}

\def\mustr{{\mu_s}}

\def\rhostr{{\rho_s}}

\def\GEn{{G_\sst{E}^n}}
\def\GEp{{G_\sst{E}^p}}
\def\GMn{{G_\sst{M}^n}}

\def\lamd{{\lambda_\sst{D}^\sst{V}}}
\def\lamn{{\lambda_n}}
\def\lams{{\lambda_\sst{E}^{(s)}}}

\def\Gdip{{G_\sst{D}^\sst{V}}}

\def\RVp{{R_\sst{V}^p}}
\def\RVn{{R_\sst{V}^n}}

\def\rbra#1{{\langle#1\parallel}}
\def\rket#1{{\parallel#1\rangle}}

\def\evec{{\vec e}}

\def\mustr{{\mu_s}}

\def\qw{{Q_\sst{W}}}
\def\alrc{{\alr(^{12}{\rm C})}}
\def\qwcs{{\qw(^{133}{\rm Cs})}}
\def\alrep{{\alr(\evec p)}}
\def\xivp{{\xi_\sst{V}^p}}
\def\xivn{{\xi_\sst{V}^n}}

\def\PRC#1{{\it Phys. Rev.} {\bf C#1} }
\def\PRD#1{{\it Phys. Rev.} {\bf D#1} }
\def\PRL#1{{\it Phys. Rev. Lett.} {\bf #1} }
\def\NPA#1{{\it Nucl. Phys.} {\bf A#1} }
\def\NPB#1{{\it Nucl. Phys.} {\bf B#1} }
\def\AoP#1{{\it Ann. of Phys.} {\bf #1} }

\def\etal{{\it et al.}}
\def\ltil{{\tilde L}}

\hfuzz=50pt

\def\nuc#1#2{{^{#1}\hbox{\rm #2}}}
\def\rviez{{R_\sst{V}^{I=0}}}
\def\rvieo{{R_\sst{V}^{I=1}}}

\def\GEIEZ{{G_\sst{E}^{I=0}}}

\def\xiviez{{\xi_\sst{V}^{I=0}}}
\def\xivieo{{\xi_\sst{V}^{I=1}}}

\vsize=7.5in
\hsize=5.6in
\magnification=1200
\tolerance 10000

\baselineskip 12pt plus 1pt minus 1pt
\pageno=0
\centerline{\bf NON-STANDARD PHYSICS AND NUCLEON STRANGENESS}
\smallskip
\centerline{{\bf IN LOW-ENERGY PV ELECTRON SCATTERING}\footnote{*}{This
work is supported in part by funds
provided by the U. S. Department of Energy (D.O.E.) under contract
\#DE-AC02-76ER03069.}}
\vskip 24pt
\centerline{M. J. Musolf\footnote{**}{Present address: Department
of Physics, Old Dominion University, Norfolk, VA 23529 and
Physics Division MS 12H, Continuous Electron Beam Accelerator Facility,
Newport News, VA  23606.} and T. W. Donnelly}
\vskip 12pt
\centerline{\it Center for Theoretical Physics}
\centerline{\it Laboratory for Nuclear Science}
\centerline{\it and Department of Physics}
\centerline{\it Massachusetts Institute of Technology}
\centerline{\it Cambridge, Massachusetts\ \ 02139\ \ \ U.S.A.}
\vskip 1.5in
\centerline{Submitted to: {\it Zeitschrift f\"ur Physik C}}
\vfill
\noindent CTP\#2149\hfill September 1992
\eject
\baselineskip 24pt plus 2pt minus 2pt
\centerline{\bf ABSTRACT}
\medskip
        Contributions from physics beyond the Standard Model, strange
quarks in the nucleon, and nuclear structure effects to the left-right
asymmetry measured in parity-violating (PV) electron scattering from
$\nuc{12}{C}$ and the proton are discussed. It is shown how lack of knowledge
of the distribution of strange quarks in the nucleon,  as well as theoretical
uncertainties associated with higher-order dispersion amplitudes and nuclear
isospin-mixing,  enter the extraction of new limits on the electroweak
parameters $S$ and $T$ from these PV observables. It is found that a series
of elastic PV electron scattering measurements using $^{4}$He could
significantly constrain the $s$-quark electric form factor if other theoretical
issues are resolved. Such constraints would reduce the associated form
factor uncertainty in the carbon and proton asymmetries below a level
needed to permit extraction of interesting low-energy constraints on
$S$ and $T$ from these observables. For comparison, the much smaller
scale of $s$-quark contributions to the weak charge measured in atomic PV
is quantified. It is likely that only in the case of heavy muonic atoms
could nucleon strangeness enter the weak charge at an observable level.

\medskip
\vfill
\eject

\medskip
\nobreak

\noindent{\bf 1. Introduction}

        It has recently been suggested that measurements of the \lq\lq
left-right" helicity difference asymmetry ($\alr$)
in parity-violating (PV) elastic electron scattering from
$^{12}$C nuclei and of the weak charge ($\qw$)
in atomic PV experiments using
$^{133}$Cs are potentially sensitive to certain extensions of the Standard
Model
at a significant level.$^{1}$ In particular, these observables
carry a non-negligible dependence on the so-called
$S$-parameter characterizing extensions of the Standard Model which involve
degenerate multiplets of heavy fermions.$^{2}$ It is argued that a
1\% measurement of $\alr(^{12}{\rm C})$ or a 0.7\% determination
of $\qw(^{133}{\rm Cs})$ (equivalent to a 1\% determination of the
weak mixing angle) would constrain $S$ to $\vert\delta S
\vert \leq 0.6$, a significant improvement over the present limit of
$\delta S = \pm 2.0\ (\hbox{\rm exp't})\ \pm 1.1\ (\hbox{\rm th'y})$
obtained from $\qwcs$.$^{1}$ The
level of systematic precision achieved in the recently completed MIT-Bates
measurement of $\alrc$,$^{3}$ along with prospects for improving statistical
precision with longer run times at CEBAF or MIT-Bates, suggest that a 1\%
$\alrc$ measurement could be feasible in the foreseeable future.
Similarly, improvements in atomic structure calculations$^{4}$ have reduced
the theoretical error in $\qwcs$ to roughly  1\%, and the prospects for pushing
the experimental uncertainty below this level also appear promising.$^{5}$
If such high-precision, low-energy measurements were achieved, the resultant
constraints on non-standard physics would complement those obtainable from
measurements in the high-energy sector. The latter are
generally equally sensitive to both $S$ and  the $T$-parameter, where the
latter characterizes standard model extensions involving non-degenerate heavy
multiplets.$^{1,\ 2}$

        In this work, we point out the presence of terms in $\alrc$
not considered in Ref. [1] involving nucleon and nuclear structure
physics which must be experimentally and/or theoretically constrained
in order to achieve the limits on $S$ suggested above.
Specifically, we consider contributions involving the distribution
of strange quarks in the nucleon, multi-boson \lq\lq dispersion corrections"
to tree-level electromagnetic (EM) and weak neutral current (NC) amplitudes,
and
isospin impurities in the nuclear ground state. We show that lack of knowledge
of $\rhostr$, the dimensionless mean square \lq\lq strangeness radius",
introduces uncertainties into $\alrc$ at a potentially problematic level. We
further show how a series of two
measurements of $\alr$ for elastic scattering from
$^{4}$He could constrain $\rhostr$ sufficiently to reduce the associated
uncertainty in $\alrc$ to below 1\% . In addition, we observe that an improved
theoretical understanding of dispersion corrections and isospin impurities
for scattering from $(J^\pi, I)=(0^+,0)$ nuclei is needed in order both
to determine $\rhostr$ at an interesting level and to constrain $S$ to
the level suggested in Ref. [1]. For comparison, we also discuss briefly
the interplay of nucleon strangeness and non-standard physics in PV elastic
$\evec p$ scattering and atomic PV. In the former instance, a 10\%
determination of $\alrep$ at forward-angles could yield
low-energy constraints on $S$ and $T$ complementary to those obtained from
either atomic PV or $\alrc$, if the strangeness radius were constrained
to the same level as appears possible with
the aforementioned series of $^{4}$He measurements. A determination of
$\rhostr$ with PV $\evec p$ scattering alone would not be sufficient for
this purpose. In contrast, the impact of strangeness on the interpretation
of $\qwcs$ is significantly smaller, down by at least an order of magnitude
from the dominant atomic theory uncertainties. Only in the case of PV
experiments with heavy muonic atoms might $\rhostr$ enter at a potentially
observable level. Other prospective PV electron scattering experiments --
such as elastic scattering from the deuteron or quasielastic scattering --
are discussed elsewhere.$^{6 - 9}$

\noindent{\bf 2. Hadronic neutral current, new physics and strangeness}

        The low-energy PV observables of interest here are dominated by
the charge ($\mu=0$) component of the hadronic vector NC. In terms of
quark fields, the nuclear
vector NC operator may be written in terms of the isoscalar and
isovector EM currents and a strange quark current:$^{6}$
$$
J_\mu^\sst{NC}=\xivieo J_\mu^\sst{EM}(I=1)+\sqrt{3}\xiviez J_\mu^\sst{EM}(I=0)
+\xivz V_\mu^{(s)}\ \ ,\eqno(1)
$$
where $V_\mu^{(s)}=\sbar\gamma_\mu s$ and the $\xi_\sst{V}$'s are couplings
determined by the underlying electroweak gauge theory. In writing Eq. (1),
we have eliminated terms involving $(c,b,t)$ quarks, since their contributions
to nuclear matrix elements of $J_\mu^\sst{NC}$ are suppressed (see below).
In the minimal Standard Model, one has
$$
\eqalignno{\xivz&=-[1+\rvz]&\cr
           \sqrt{3}\xiviez&=-4\sstw[1+\rviez]&(2)\cr
           \xivieo&=2(1-2\sstw)[1+\rvieo]\ \ ,&\cr}
$$
where $\sstw$ is the weak mixing angle and the $R_\sst{V}^{(a)}$ are
higher-order corrections to tree-level electron-nucleus
NC amplitudes. In addition, one may define couplings which govern the
low-$|Q^2|$ NC charge scattering from the neutron and proton:
$$
\eqalignno{\xivp&\equiv\coeff{1}{2}[
        \sqrt{3}\xiviez+\xivieo]\>=\>(1-4\sstw)[1+\RVp]&\cr
           \xivn&\equiv\coeff{1}{2}[
        \sqrt{3}\xiviez-\xivieo]\>=\>-[1+\RVn]\ \ .&(3)\cr}
$$
At the operator level,
the $\xi_\sst{V}$'s are determined entirely in terms of couplings of the
$Z^0$ to the $(u,d,s)$ quarks, including contributions from radiative
corrections within or beyond the framework of the Standard Model,
both of which may be included in the $R_\sst{V}^{(a)}$.$^{10}$
Upon taking nuclear matrix elements of $J_\mu^\sst{NC}$, one must include
in the $R_\sst{V}^{(a)}$ additional contributions arising from strong
interactions between quarks in intermediate states.$^{10,\ 11}$ Further
contributions arising from isospin impurities in the nuclear ground state
are discussed below. Corrections owing to neglect of the $(c,b,t)$ quarks in
writing Eq. (1) have been estimated in Ref. [12] and may be included in
the $R_\sst{V}^{(a)}$ for $a=0$ and $I=0$
as $R_\sst{V}^{(a)}\rightarrow R_\sst{V}^{(a)}({\rm ewk})
-\Delta_\sst{V}$, where $\Delta_\sst{V}\sim 10^{-4}$. No such corrections
enter $\rvieo$.

        The motivation for considering PV electron scattering as a probe of
new physics may be seen, for example, by noting the $S$- and $T$-dependencies
of the $R_\sst{V}^{(a)}$. Following Ref.~[1], in which $\overline{\hbox{MS}}$
renormalization was used in computing one-loop electroweak corrections, one
has
$$
\eqalignno{\rviez({\rm new})&=0.016S-0.003T&\cr
           \rvieo({\rm new})&=-0.014S+0.017T&(4)\cr
           \RVp({\rm new})&=-0.206S+0.152T&\cr
           \RVn({\rm new})&=0.0078T\ \ \ \ \ .&\cr}
$$
Within the framework of Ref.~[1], a value of the top-quark mass differing
from 140 GeV would also generate a non-zero contribution to $T$. The
different linear combinations of $S$ and $T$ appearing in Eqs.~(4)
suggest that a combination of PV electron scattering experiments could
provide interesting low-energy constraints on these two parameters. One
such scenario is illustrated in Fig. 1,
where the constraints attainable from a 1\%
measurement of $\alrc$ and a 10\% determination of $\xivp$ from a
forward-angle $\alrep$ measurement are shown. For comparison, the
present constraints from $\qwcs$ are also shown. One expects these
constraints to be tightened by a factor of two to three with future
measurements.$^{13}$ While $\qwcs$ is
effectively independent of $T$, both $\alrc$ and the forward-angle $\evec p$
asymmetry carry a non-negligible dependence on $T$. Hence, one or both
of the latter could complement the former as a low-energy probe of new
physics.In addition, one might also consider PV electro-excitation of
the $\Delta(1232)$ resonance as a means of extracting $\rvieo$. This
quantity is relatively more sensitive to $T$ than are $\rviez$, $\rvz$, and
$\qwcs$, so that a determination of the former would further complement
any low-energy constraints attained from the latter.$^{14}$ It is unlikely,
however, that the experimental and theoretical uncertainties associated
with $\alr(N\to\Delta)$ will be reduced to the level necessary to make
such a measurement relevant as an electroweak test in the near term.$^{8,\ 14}$
Consequently, a combination of PV scattering experiments on $^{12}$C and/or
the proton, together with atomic PV, appear to hold the most promise for
placing low-energy, semileptonic constraints on new physics. Before such a
scenario  is realized, however, other hadronic physics dependent terms entering
the PV asymmetries must be analyzed. We now consider these additional
contributions, focusing first on the simplest case of $\nuc{12}{C}$.

\noindent{\bf 3. PV elastic scattering from carbon}

        In the limit that the $^{12}$C ground state is an eigenstate of
strong isospin, matrix elements of the isovector component of the current in
Eq. (1) vanish. Moreover, since this nucleus has zero spin, only monopole
matrix
elements of the charge operator contribute. In the absence of the
strange-quark term in Eq. (1), one has $\rbra{{\rm g.s.}}\rho^\sst{NC}\rket{{
\rm g.s.}}=\sqrt{3}\xiviez\rbra{{\rm g.s.}}\rho^\sst{EM}\rket{{\rm g.s.}}$, so
t
hat
$\alrc\propto\rbra{{\rm g.s.}}\rho^\sst{NC}\rket{{\rm g.s.}}/\rbra{{\rm g.s.}}
\rho^\sst{EM}\rket{{\rm g.s.}}=\sqrt{3}\xiviez$. In short, the asymmetry
becomes
independent of the nuclear physics contained in the EM and NC matrix
elements$^{15, \ 16}$
and carries a dependence only on the underlying gauge theory coupling,
$\xiviez$. Upon including the strange-quark term one has$^{6,\ 17}$
$$
\alrc\>=\>A_0 Q^2\left[4\sstw(1+\rviez) + {\GES(Q^2)\over \GEIEZ}(1+\rvz)
\right]\ \ ,\eqno(5)
$$
where $A_0=G_\mu/(4\sqrt{2}\pi\alpha)=8.99\times 10^{-5} {\rm GeV}^{-2}$,
$G_\mu$ is the Fermi constant measured in muon decay,
$Q^2=\omega^2-|\svec{q}|^2\leq 0$ is the four-momentum transfer squared, and
$\GES(Q^2)$ and $\GEIEZ(Q^2)$ are the Sachs electric form factors$^{18}$
appearing in single-nucleon
matrix elements of $V_\mu^{(s)}$ and $J_\mu^\sst{EM}(I=0)$. Note that at
the one-body level, the strangeness and EM charge density operators,
$\hat\rho^{(s)}$ and $\hat\rho^\sst{EM}(I=0)$, respectively, are identical,
apart from the single nucleon form factors which enter multiplicatively.
Consequently, any dependence on the nuclear wavefunction cancels from the
asymmetry, leaving only the ratio of form factors in the second term of
Eq.~(5). For $\rviez$ one has
$$
\rviez\>=\>\rviez(\hbox{\rm st'd})+\rviez(\hbox{\rm new})-
        \rviez(\hbox{\rm QED})+\rviez({\rm had})+\Gamma -\Delta_\sst{V}\ \ ,
\eqno(6)
$$
where $\rviez(\hbox{\rm st'd})$ are Standard Model electroweak radiative
correct
ions
to tree-level electron quark PV NC amplitudes, $\rviez(
\hbox{\rm new})$ denote
contributions from extensions of the Standard Model as in Eqs.~(4),
{ $\rviez(\hbox{\rm
QED})$ are QED radiative corrections to the EM amplitude entering the
denominator of $\alrc$ (hence, the minus sign in Eq. (6)),
$\rviez({\rm had})$ are
strong-interaction hadronic contributions to higher-order electroweak
amplitudes, $\Gamma$ is a correction due to isospin impurities in the
$^{12}$C ground state,$^{19}$ and $\Delta_\sst{V}$ is the heavy-quark
correction discussed previously. The correction $\rvz$ appearing in the
second term of Eq. (5) may be written in a similar form.
For fixed top-quark and Higgs masses, the
$\rviez(\hbox{\rm st'd})$  and $\rviez(\hbox{QED})$
can be determined unambiguously, up to hadronic
uncertainties associated with quark loops in the $Z^0-\gamma$ mixing tensor
and two boson-exchange \lq\lq box" diagrams ( see, {\it e.g.}, Ref. [11]).

        Before discussing the remaining terms in Eq.~(6), we note here an
additional feature of spin-0 nuclei which simplifies the interpretation
of the PV asymmetry. In general, when working to one-loop order, one
must also include bremsstrahlung contributions to the helicity-dependent
(-independent) cross sections entering the numerator (denominator) of $\alr$.
These contributions, although not loop corrections, enter the cross section
at the same order in $\alpha$ as one-loop amplitudes and should be formally
included in the $\rviez(\hbox{\rm st'd})$  and $\rviez(\hbox{\rm QED})$.
At low momentum transfer,
one need only consider bremsstrahlung from the scattering electron (Fig. 2),
since the target experiences very small recoil and is unlikely to radiate.
The contributions to the EM and EM-NC interference cross sections from the
amplitudes of Fig. 2 are
$$
\eqalignno{d\sigma^{\rm brem}_\sst{EM}&\propto |M_a+M_b|^2 = M_a M_a^*+
                M_a M_b^* + M_b M_a^* + M_b M_b^* &(7{\rm a})\cr
           d\sigma^{\rm brem}_\sst{INT}&\propto M_a M_c^* + M_a M_d^*+
                M_b M_c^*+ M_b M_d^* + \hbox{\rm c.c.}\ \ ,&(7{\rm b})\cr}
$$
where the $M_{i}$ are the amplitudes associated with the diagrams in Fig. 2.
For simplicity, we consider only the first terms on the right side of
Eqs.~(7). The arguments for the remaining terms are similar.  For these
terms one has
$$
\eqalignno{M_a M_a^* &= {(4\pi\alpha)^3\over Q^4} \ltil_{\mu\nu}^\sst{EM}
        W^{\mu\nu}_\sst{EM}&(8{\rm a})\cr
           M_a M_c^* &= -{(4\pi\alpha)^2\over Q^2}{G_\mu\over 2\sqrt{2}}
        \ltil_{\mu\nu}^\sst{INT} W^{\mu\nu}_\sst{INT}\ \ ,&(8{\rm b})\cr}
$$
where the $W^{\mu\nu}$ are hadronic tensors formed from products of the
hadronic electromagnetic and weak neutral currents,
and where the $\ltil_{\mu\nu}$ are the corresponding
tensors formed from the leptonic side of the diagrams in Fig. 2.
The $W^{\mu\nu}$ are identical to the tree-level hadronic tensors, since
the only differences between the diagrams of Fig. 2 and the tree-level
graphs involve the lepton line. For the leptonic tensors, one has after
averaging over initial and summing over final states$^{20}$
$$
\eqalignno{\ltil_{\mu\nu}^\sst{EM}&=\coeff{1}{2}[(K'+q)^2-\mes]^{-2}
        \hbox{\rm Tr}\Bigl\{\gamma_\lambda (\notKp+\notq+\me)\gamma_\mu
        (1+\gamma_5\nots)&(9{\rm a})\cr
         &\times(\notK+\me)\gamma_\nu(\notKp+\notq+\me)\gamma_\sigma
        (\notK+\me)\Bigr\}\varepsilon^\lambda\varepsilon^{\sigma}&\cr
        &&\cr
        \ltil_{\mu\nu}^\sst{INT}&=\coeff{1}{2}[(K'+q)^2-\mes]^{-2}
        \hbox{\rm Tr}\Bigl\{\gamma_\lambda (\notKp+\notq+\me)\gamma_\mu&\cr
        &\times(\gve+\gae\gamma_5)(1+\gamma_5\nots)&(9{\rm b})\cr
        &(\notK+\me)\gamma_\nu(\notKp+\notq+\me)\gamma_\sigma
        (\notK+\me)\Bigr\}\varepsilon^\lambda\varepsilon^{\sigma}\ \ ,&\cr}
$$
where $K_\mu$ ($K_\mu^{\prime}$) are the initial (final) electron momenta,
$q_\mu$ is the momentum of the outgoing photon having polarization
$\varepsilon_\mu$, $s_\mu$ is the initial electron spin, and $\gve$ ($\gae$)
are the vector (axial vector) NC couplings of the electron.

        Taking the electron and radiated photon on-shell ($K^2=K^{\prime\, 2}
=\mes$, $q^2=0$) and working in the extreme relativistic limit ($E_e/\me >> 1$)
for which $s_\mu\to (h/\me)K_\mu$, with $h$ being the electron helicity, one
has
$$
\eqalignno{M_a M_a^*&={(4\pi\alpha)^3\over Q^4}{1\over 2}\biggl({1\over 2 K'
        \cdot
        q}\biggr)^2&\cr
        &\times\hbox{\rm Tr}\Bigl\{\gamma_\nu(\notKp+\notq)\gamma_\sigma
        \notKp\gamma_\lambda(\notKp+\notq)\gamma_\mu\notK\Bigr\}
        \varepsilon^\lambda\varepsilon^{\sigma}W^{\mu\nu}_\sst{EM}&(10{\rm
        a})\cr
        &&\cr
        M_a M_c^*&=-{(4\pi\alpha)^2\over Q^2}{G_\mu\over 2\sqrt{2}}
        {h\over 2}\biggl({1\over 2 K'\cdot
        q}\biggr)^2&\cr
        &\times\biggl[-\gve\hbox{\rm Tr}
        \Bigl\{\gamma_\nu(\notKp+\notq)\gamma_\sigma
        \notKp\gamma_\lambda(\notKp+\notq)\gamma_\mu\notK\gamma_5\Bigr\}&
        (10{\rm b})\cr
        & +\gae\hbox{\rm Tr}
        \Bigl\{\gamma_\nu(\notKp+\notq)\gamma_\sigma
        \notKp\gamma_\lambda(\notKp+\notq)\gamma_\mu\notK\Bigr\}\biggr]
        \varepsilon^\lambda\varepsilon^{\sigma}W^{\mu\nu}_\sst{INT}\ \ .
        &\cr}
$$

        For elastic scattering from spin-0 nuclei, only the $\mu=\nu=0$
components of the $W^{\mu\nu}$ are non-vanishing. Since the trace multiplying
$\gve$ in Eq.~(10b) is anti-symmetric in $\mu$ and $\nu$, this term
does not contribute. Adding Eqs.~(10) to the absolute squares of the
corresponding tree-level amplitudes leads to
$$
\eqalignno{d\sigma_\sst{EM}^{\rm tree}+d\sigma_\sst{EM}^{\rm brem}&\sim
        {1\over 2}{(4\pi\alpha)^2\over Q^4}\biggl[\hbox{\rm Tr}\Bigl\{\gamma_0
        \notKp\gamma_0\notK\Bigr\}\> +&\cr
        &(4\pi\alpha)\hbox{\rm Tr}\Bigl\{\gamma_0(\notKp+\notq)
        \gamma_\sigma\notKp\gamma_\lambda(\notKp+\notq)\gamma_0\notK\Bigr\}
        \varepsilon^\lambda\varepsilon^{\sigma}\biggr]
        W^{00}_\sst{EM}&(11{\rm a})\cr
        &&\cr
        d\sigma_\sst{INT}^{\rm tree}+d\sigma_\sst{INT}^{\rm brem}&\sim
        -{h\over 2}{(4\pi\alpha)\over Q^2}{G_\mu\over 2\sqrt{2}}\gae
        \biggl[\hbox{\rm Tr}\Bigl\{\gamma_0\notKp\gamma_0\notK\Bigr\}\> +&\cr
        &(4\pi\alpha)\hbox{\rm Tr}\Bigl\{\gamma_0(\notKp+\notq)
        \gamma_\sigma\notKp\gamma_\lambda(\notKp+\notq)\gamma_0\notK\Bigr\}
        \varepsilon^\lambda\varepsilon^{\sigma}\biggr]
        W^{00}_\sst{INT}\ \ .&(11{\rm b})\cr}
$$
Since $\alr=(d\sigma_\sst{INT}^+ - d\sigma_\sst{INT}^-)/d\sigma_\sst{EM}$,
and since the quantities inside the square brackets in Eqs.~(11a) and (11b)
are identical, they cancel from the asymmetry. It is straightforward to
show that this cancellation occurs even when the remaining terms in Eqs.~(7)
are included. In short, the bremsstrahlung contributions drop out entirely
from $\alr$, leaving the expression of Eq.~(5) unchanged. One could, of course,
attempt to be more rigorous and integrate bremsstrahlung cross sections over
the detector acceptances, {\it etc}. In doing so, however, one would only
modify
the form of the expressions inside the square brackets
in Eqs.~(11) and not change the
fact that they are identical in the two equations. The cancellation of
bremsstrahlung contributions to the asymmetry would still obtain in this case.
We note that this result does not carry over to nuclei having spin $> 0$. In
the
latter case, $\alr$ receives contributions from the leptonic vector NC
(first term on the right side of Eq.~(10)). There exists no term in
$d\sigma_\sst{EM}^{\rm brem}$ to cancel the corresponding contribution from
$d\sigma_\sst{INT}^{\rm brem}$.

        Returning to the remaining terms in Eq.~(6), we emphasize that
in contrast to the first three terms, the remaining terms are theoretically
uncertain, due to the present lack of tractable methods for calculating
low-energy strong interaction dynamics
from first principles in QCD. Of particular concern are
multi-boson-exchange dispersion contributions to $\rviez({\rm had})$,
such as those generated
by the diagrams of Fig. 3. We note that neither the $\GES$-term of Eq. (5)
nor the nuclear, many-body contributions to the dispersion corrections were
included in the discussion of Ref. [1].

        We first consider the impact of strangeness on the extraction of
$S$ from $\alrc$. To that end, we employ an \lq\lq extended" Galster
parameterization$^{21}$ for
the single-nucleon form factors appearing in Eq. (5):
$\GEIEZ=\coeff{1}{2}\left[\GEp+\GEn\right]$, $\GEp=\Gdip$, $\GEn=-\mu_n\tau
\Gdip\xi_n$, $\GES=\rhostr\tau\Gdip\xi_s$,
where $\mu_n$ is the neutron magnetic moment, $\tau=-Q^2/4\mns$,
$\Gdip=(1+\lamd\tau)^{-2}$ is the standard dipole form factor
appearing in nucleon form factors, and
$\xi_{n,s}=(1+\lambda_\sst{E}^{(n,s)}\tau)^{-1}$ allow for more rapid
high-$|Q^2|$ fall-off than that given by the dipole form factor. From
parity-conserving electron scattering, one has $\lamd\approx4.97$ and
$\lamn\approx5.6$.$^{21}$ It is possible that $\GES$ falls off more rapidly
at high-$|Q^2|$ than the $1/Q^4$ behavior exhibited by this parameterization,
but for the momentum transfers of interest here,$^{22}$
this choice is sufficient. The parameters $\rhostr$ and $\lams$ characterize
the
low- and moderate-$|Q^2|$ behavior, respectively, of $\GES$ and are presently
un-constrained. Because the nucleon has no net strangeness, $\GES$ must
vanish at $Q^2=0=\tau$. Hence, like $\GEn$, which also must vanish at the
photon point, the $\GES$ carries a linear dependence on $|Q^2|$ near
the photon point. While no experimental information on $\GES$ exists,
theoretical predictions for the mean-square strangeness radius (of which
$\rhostr$ is a dimensionless version) have been made using different
models.$^{21 - 25,\ 14}$ Since these models generally predict qualitatively
different behaviors of $\GES$ at moderate-$|Q^2|$, we choose the simple and
convenient Galster-like parameterization in which
variations in this moderate-$|Q^2|$ behavior
are characterized by a single parameter $\lams$ to be constrained by
experiment.

        Under these choices, the strange-quark term in Eq. (5) induces a
fractional shift in the $\alrc$ asymmetry given by
$$
{\Delta\alr\over\alr}={\rhostr\tau\xi_s\over 2\sstw[1-\mu_n\tau\xi_n]}\eqno(12)
$$
neglecting $\rvz$. Taking the average value for $\rhostr$ predicted in Ref.
[22], choosing $\lams=\lamn$, and working at the kinematics of the recent
MIT-Bates $\alrc$ measurement ($\tau\approx 0.007$), Eq. (12) indicates about
a -3\% shift in $\alrc$. Any {\it uncertainty} in $\GES$ on this scale would
weaken by a factor of three the limits on $S$ predicted in Ref. [1].

        From the standpoint of reducing the uncertainty in $\alrc$ Standard
Model tests, as well as that of learning about the distribution of strange
quarks in the proton, it is clearly desirable to constrain $\GES$ as tightly
as possible. To that end, a combination of two measurements of
$\alr$ on a $(0^+,0)$ target could constrain $\GES$ sufficiently to reduce
the $\GES$-induced error in a subsequent determination of $S$ from $\alrc$
to below $|\delta S|=0.6$. For this purpose, we consider $^{4}$He rather than
$^
{12}$C.
The statistical precision, $\delta\alr/\alr$, achievable for either nucleus
goes as ${\cal F}^{-1/2}$, where the figure of merit ${\cal F}=\sigma \alr^2$,
with $\sigma$ being the EM cross section.$^{6}$ For both
nuclei, $\delta\alr/\alr$ displays a succession of local minima as a function
of $|Q^2|$, corresponding to successive local maxima in the cross section.
Since
the relative sensitivity of $\GES$ to $\rhostr$ and $\lams$ changes with
$|Q^2|$, a measurements of $\alr(0^+,0)$ in the vicinity of different local
minima in $\delta\alr/\alr$ would impose somewhat different joint constraints
on $\rhostr$ and $\lams$. The EM cross section falls off more gently with
$|Q^2|$ for $^{4}$He than for $^{12}$C, so that for the former, the first
two $\delta\alr/\alr$ minima are more widely separated in $|Q^2|$ than for
the latter. Consequently, the constraints on $\GES$ obtainable with
two measurements carried out, respectively, at the first two $\delta\alr/\alr$
minima on $^{4}$He could be more restrictive than with a similar series
involving $^{12}$C.

        To complete this analysis, we consider a combination of two such
$\alr(^4{\rm He})$ experiments carried out roughly
under conditions that are representative of what could be achievable with a
moderate solid angle detector at
CEBAF: luminosity ${\cal L}=5\times 10^{38} {\rm cm}^{-2}{\rm
s}^{-1}$, scattering angle $\theta = 10^\circ$, solid angle $\Delta\Omega =
0.01$ steradians, beam polarization $P_e = 100 \%$, and run time $T=1000$
hours.$^{26}$\ The constraints resulting from these two prospective
measurements are shown in Fig. 4. Since nothing at present is know
experimentally about $\GES$, we assume two different models for illustrative
purposes: (A) $(|\rhostr|, \lams) = (0, \lamn)$ and (B) $(|\rhostr|, \lams) =
(2, \lamn)$. The value of $|\rhostr|$ in
model (B) corresponds roughly to the average prediction of Ref. [22].
{}From these results, we find that for model (B), the uncertainty
remaining in $\GES$ after the series of $^{4}$He measurements would be
sufficiently small to keep the associated error in a lower-$|Q^2|$ Standard
Model test with either $^{12}$C or $^{4}$He below 1\%.
In the case of model (A), even though $\lams$ is not constrained, the
lower-$|Q^2|$ measurement appears to keep the $\GES$-induced error in a
$(0^+,0)$ Standard Model test below 1\% , independent of the value of $\lams$.

        Before such $^{4}$He constraints could be attained or a 1\% Standard
Model test performed,
ambiguities associated with dispersion corrections in $\rviez({\rm
had})$ and with the isospin-mixing parameter $\Gamma$ must be resolved. Turning
first to the former,  we focus on {\it nuclear} many-body contributions
to the amplitudes associated with Fig. 3.  Since
$\alr(0^+,0)\sim M^\sst{PV}_\sst{NC}(I=0)/
M^\sst{PC}_\sst{EM}(I=0)$, where $M^\sst{PV}_\sst{NC}(I=0)$
($M^\sst{PC}_\sst{EM}(I=0)$) are the isoscalar parity-violating (-conserving)
scattering amplitudes, and since the dispersion corrections enter as ${\cal
O}(\alpha)$ corrections to the tree-level amplitudes, one has
$\rviez({\rm disp})\sim R_\sst{V}^{VV'}(I=0)-R_\sst{V}^{\gamma\gamma}(I=0)$,
where $R_\sst{V}^{VV'}$ is a dispersion correction to the tree-level
$Z^0$-exchange amplitude involving one or more heavy vector bosons and
$R_\sst{V}^{\gamma\gamma}$ is the two-photon correction to the isoscalar
electromagnetic amplitude. Although one might na\"\i vely hope for some
cancellation between these two corrections, the different
$Q^2$-dependences carried by each makes such a possibility unlikely. Whereas
$R_\sst{V}^{\gamma\gamma}\to 0$ as $|Q^2|\to 0$, since the tree-level
EM amplitude has a pole at $Q^2=0$, $R_\sst{V}^{VV'}$ need not vanish in
this limit since the tree-level NC amplitude has a pole at $Q^2=\mzs$.

Generally speaking, one expects the scale of hadronic contributions to
$\rviez(\hbox{disp})$ to be of ${\cal O}(\alpha/4\pi)$. Indeed, theoretical
estimates of such contributions to the 2-$\gamma$,  PC, $ep$ scattering
amplitud
e
indicate that $R_\sst{V}^{\gamma\gamma}(ep) \lapp 1\%$ at intermediate
energies.$^{27,\ 28}$ However,
experimental information on $R_\sst{V}^{\gamma\gamma}$ suggests
that the dispersion corrections for scattering from {\it nuclei} can be
significantly larger than the one-body ($ep$) scale. Results from
the recent MIT-Bates measurement of $R_\sst{V}^{\gamma\gamma}(I=0)$ for
$^{12}$C show that this correction could be as large as 20\% in the
first diffraction minimum and several percent in the regions outside the
minimum where a $(0^+,0)$ Standard Model test or $\GES$-determination might
be undertaken.$^{29}$ In the latter regions, the experimental error in
$R_\sst{V}^{\gamma\gamma}(I=0)$ is of the same order as the correction itself,
and the overall level of agreement between these results and
theoretical calculations$^{30}$ is rather poor. In short, experimentally
and theoretically uncertain many-body effects appear to enhance the
scale of $R_\sst{V}^{\gamma\gamma}(I=0)$ to a level which is important for the
interpretation of $\alr(0^+,0)$.

In the case of PV amplitudes, no experimental information
exists on $R_\sst{V}^{VV'}(I=0)$. It is unlikely that this quantity
will be measured directly,
so that one must rely on nuclear model-dependent theoretical
estimates of its scale. Of particular concern is the $Z^0-\gamma$
dispersion amplitude which,
for elementary $e-q$ scattering, contains  logarithms
involving the ratios $|\mzs/s|$ and $|\mzs/u|$,
where the scale of the invariant variables $s$ and $u$
is set by the incoming electron momentum and the typical momentum of the
quark bound in the target nucleus.$^{10}$
This logarithmic scale mismatch suggests
that contributions from low-energy intermediate states involving
hard-to-calculate hadronic collective excitations ({\it e.g.}, the nuclear
giant resonance) could be important. Given the scale of the
$R_\sst{V}^{\gamma\gamma}(I=0)$ results, the discrepancy with theory, and the
need for a theoretical estimate of $R_\sst{V}^{Z\gamma}(I=0)$,
significant progress in theoretical understanding of
many-body contributions to the  dispersion corrections is
needed in order to keep the corresponding uncertainty in $\alr(0^+,0)$ below
one percent.

The quantity $\Gamma (q)$ $(q \equiv |\ {\vec q}\ |)$
in Eq.~(6) has been introduced to take into
account the fact that nuclei such as $^4$He and $^{12}$C are not exact
eigenstates of strong isospin with $I=0$. Since, the EM interaction does
not conserve isospin, one expects states having
$I\not= 0$ to be present as small [{\cal O}($\alpha$)] components
in the nuclear ground states.
For nuclei whose major configurations involve either
the $1s$ shell ($^4$He) or the $1p$ shell ($^{12}$C) the isospin-mixing
correction $\Gamma(q)$
is likely to be quite small at low momentum transfer
($|\Gamma| \lapp 1\%$).$^{19}$
This special situation arises because of the difficulty of
supporting isovector breathing modes in the relevant nuclear model spaces;
since  primarily a single type of radial wave function plays a role, radial
excitations are suppressed.

        We emphasize that this conclusion need not apply
to spin-0 nuclei beyond the $1s-1p$ shell.
For nuclei in the $2s-1d$ shell, for example, one has wave functions which
display different radial distributions ({\it viz.,\/} $2s$ and $1d$),
making it possible to have important isovector breathing-mode
admixtures introduced into the nuclear ground states.
For nuclei beyond $^{40}$Ca an additional issue arises. Since
in this region the stable 0$^+$ nuclei have $N>Z$ and, thus,
$I\not= 0$ from
the outset, both isoscalar and isovector matrix elements of the
monopole operators enter (even in the absence of isospin-mixing). In this case,
isospin-mixing effects appear in two ways:
(1) several  eigenstates of isospin
can mix to form the physical states (as above) and (2) the mean
fields in which the protons and neutrons in the nucleus move may be slightly
different.  This latter effect was explored in Ref.~[19], where it was
found that $\alr$ for elastic scattering from 0$^+$ $N>Z$ nuclei
is rather sensitive to the difference between $R_p$ and $R_n$, the radii of
the proton and neutron distributions in the nuclear ground state, respectively.
The reason for this sensitivity is that $|\xivn| >> |\xivp|$, making the NC
\lq\lq charge" densities for the neutron and proton roughly comparable in
magnitude.

These observations imply that the extraction of interesting constraints
on $S$, $T$, and $\GES$ from measurements of $\alr$ for spin-0 nuclei
in this region is likely to be more difficult than for spin-0 nuclei
in the $1s$-$1p$ shell. On the other hand, such measurements could provide
a new window on certain aspects of nuclear structure. Since the EM charge
radius
can be determined quite precisely using, {\it e.g.},
parity-conserving (PC) electron scattering,
a measurement of $\alr$ would provide a way to determine $R_n$.
A 1\% determination of
$R_n$ appears to be achievable.  For a nucleus such as
$^{133}$Cs, with its importance for atomic PV, it may prove useful
to employ electron scattering to explore some of these issues. The charge
and neutron distributions could be studied, thereby helping to reduce
$R_n$ uncertainties appearing in $\qwcs$ (see Eq.~(16) below), and
some indication concerning the
degree of isospin-mixing [Eq.~(16)] could be obtained.

\noindent{\bf 4. PV elastic scattering from the proton}

        As illustrated in Fig. 1, $\alrep$ carries a stronger dependence on
$T$ than either $\qwcs$ or $\alrc$, so that a measurement of the former, in
combination of one or both of the latter, could provide an interesting set
of low-energy constraints on $S$ and $T$. Na\"\i vely, one might expect the
interpretation of $\alrep$ to be simpler than that of $\alrc$, since one has
no many-body nuclear effects to take into account. However, the spin and
isospin quantum numbers of the proton allow for the presence of several form
factors in $\alrep$ not appearing in the $\nuc{12}{C}$ asymmetry, with the
result that the interpretation of PV $\evec p$ scattering is in some respects
more involved than that of elastic scattering from $(0^+,0)$ nuclei. A detailed
discussion of PV elastic $\evec p$ scattering can be found in Refs.~[6, 31,
32], and we focus here solely on scattering in the forward direction.

        At low momentum transfer and in the forward direction, the $\evec p$
asymmetry has the form$^{6}$
$$
\alrep\>\approx\>
a_o\tau\Bigl[\xivp-\Bigl\{\GEn+\GES+\tau\mu_p(\GMn+\GMS)\Bigr\}\Bigr]+
\hcal{O}(\tau^2)\ \ ,\eqno(13)
$$
where $a_o\approx 3\times 10^{-4}$. The first term on the right side of
Eq.~(13) (containing $\xivp$)
is nominally independent of hadronic physics for essentially
the same reasons as is the first term in the carbon asymmetry of Eq.~(5).
The terms contained inside the curly brackets
all enter at $\hcal{O}(\tau)$, since
both $\GEn$ and $\GES$ vanish at the photon point. From Eq.~(13) one sees
immediately the additional complexity of the proton asymmetry in comparison
with that of carbon. The neutron EM form factors appear in $\alrep$, since
the isovector and isoscalar EM currents enter the hadronic neutral current
(Eq.~(1)) with different weightings than in the hadronic EM current. The
presence of these form factors introduces one source of uncertainty not present
at the same level in $\alrc$. In addition, both the electric and magnetic
strangeness form factors contribute at $\hcal{O}(\tau)$, and their presence
also complicates the interpretation of the asymmetry.

        As in the case of $\alrc$, the $\tau$-dependence of the terms
in Eq.~(13) suggests a two-fold strategy of measurements: (a) a very low-$\tau$
measurement to determine $\xivp$, with an eye to obtaining the constraints
indicated in Fig. 1, and (b) a moderate-$\tau$ measurement aimed at
constraining the linear combination of form factors appearing in the second
term of Eq.~(13). The second of these measurements could be of interest for
a number of reasons: to extract limits on the strangeness form factors, to
constrain $\GES$ for purposes of interpreting $\alrc$ as a Standard Model
test, or to constrain this term for the same purpose but with a very low-$\tau$
$\alrep$ measurement. Considering first scenario (a), we note that it is
not possible to perform a Standard Model test at arbitrarily low-$\tau$,
since the statistical uncertainty {\it increases} for decreasing momentum
transfer. For purposes of illustration, then, we analyze a prospective
measurement at the limits of $\tau$ and forward scattering angle expected to
be achievable at CEBAF Hall C. In order to achieve the 10\% statistical
uncertainty needed for the constraints in Fig. 1, a 1000 hour experiment
would be needed, assuming 100\% beam polarization. Under these conditions,
the impact of form factor uncertainties on a determination of $\xivp$ is
non-negligible. The dominant uncertainty is introduced by $\GES$. An
uncertainty in the strangeness radius of $\delta\rhostr=\pm 2$ (corresponding
to the magnitude of the prediction in Ref.~[22]) would induce nearly a
30\% uncertainty in the extracted value of $\xivp$, a factor of three
greater than the uncertainty assumed in Fig. 1. Similarly, an uncertainty
in the value of $\mustr$ of $\pm 0.3$, also corresponding to the magnitude
of the prediction in Ref.~[22], would generate roughly a 20\% error in
$\xivp$.

        These statements point to the need for better constraints on the
strangeness form factors if an interesting Standard Model test is to be
performed with PV $\evec p$ scattering. Turning, then, to strategy (b), we
consider the constraints one might place on these form factors with a
moderate-$\tau$ $\alrep$ measurement. The difficulty here is that it is not
possible to separate the form factors with $\evec p$ scattering alone. As
discussed in Ref.~[6], a \lq\lq perfect" backward-angle $\alrep$ measurement
(0\% experimental error) might ultimately allow a determination of $\mustr$
with an error of $\pm 0.12$, thereby reducing the $\mustr$-induced uncertainty
in a forward-angle Standard Model test below a problematic level. A subsequent
determination of the second term in Eq.~(13) might then allow a determination
of $\GES$. We show in Fig. 4 the constraints in $(\rhostr,\lams)$ space
such a measurement might achieve, assuming experimental conditions similar
to those of recent CEBAF proposals.$^{33 - 35}$ We note that these constraints
would not be sufficient to permit either a 10\% determination of $\xivp$
from a low-$\tau$ $\alrep$ measurement or a 1\% Standard Model test with
elastic scattering from $\nuc{12}{C}$. In the former case, the $\GES$-induced
uncertainty in $\xivp$ would still be on the order of 20\% . In fact, as
Fig. 4 illustrates, it appears that
a series of $\alr(\nuc{4}{He})$ measurements could place
far more stringent limits on $\GES$ than appears possible with PV $\evec p$
scattering alone. Indeed, these limits would be sufficient to permit one
to probe new physics with both $\alrep$ and $\alrc$ at the level assumed in
Fig. 1.

\noindent{\bf 5. Atomic PV}

        One should expect the impact of form factor uncertainties on
the interpretation of $\qw$ to be considerably smaller than for electron
scattering asymmetries, due to the very small effective momentum-transfer
associated with the interaction of an atomic electron with the nucleus. Below,
we quantify this statement with regard to the strangeness form factors, and
note that only in the case of PV experiments with heavy muonic atoms might
nucleon strangeness contribute at an observable level. To that end,
consider the PV atomic hamiltonian
which induces mixing of opposite-parity atomic states and leads to the presence
of $\qw$-dependent atomic PV observables:
$$
\Hhat^{\rm atom}_\sst{PV}={G_\mu\over 2\sqrt{2}}\int d^3x {\hat\psi^{\dag}_e}(
\xv)\gamma_5{\hat\psi_e}(\xv) \rho^\sst{NC}(\xv) + \cdots\ \ ,\eqno(14)
$$
where ${\hat\psi_e}(\xv)$ is the electron field and $\rho^\sst{NC}(\xv)$
is the Fourier Transform of $\rho^\sst{NC}(\qv)$, the matrix element of the
charge component of Eq. (1). For simplicity, we have omitted terms involving
the spatial components of the nuclear vector NC as well as the nuclear axial
vector NC. For a heavy atom, the leading term in
Eq. (14) is significantly enhanced relative
to the remaining terms by the coherent behavior of the nuclear charge operator.
Consequently, one typically ignores the contribution from all magnetic form
factors.
Following Ref.~[36], we write the matrix element of the leading term in
$\Hhat^{\rm atom}_\sst{PV}$
between atomic $S_{1/2}$ and $P_{1/2}$ states in the form $\bra{P}
{\hat\psi^{\dag}_e}
(\xv)\gamma_5{\hat\psi_e}(\xv)\ket{S}={\cal N} C_{sp}(Z) f(x)$, where ${\cal
N}$ is a known overall normalization, $C_{sp}(Z)$ is an atomic
structure-depende
nt
function, and $f(x)=1-\coeff{1}{2}(x/x_o)^2+\cdots$ gives the
spatial-dependence of the electron axial charge density. In a simple model
where a charge-$Z$ nucleus is taken as a sphere of constant electric charge
density out to radius $R$, one has $x_o=R/Z\alpha$ neglecting small corrections
involving the electron mass. In this case, atomic matrix elements of Eq. (14)
become
$$\bra{P}\Hhat^{\rm atom}_\sst{PV}\ket{S} = {G_\mu\over 2\sqrt{2}}{\cal N}
C_{sp}(Z)\left[Q_\sst{W}^{(0)}+\Delta Q_\sst{W}^{(n,\ p)}+\Delta
Q_\sst{W}^{(s)}
+\Delta Q_\sst{W}^{(I)}\right] +\cdots\ \ ,\eqno(15)
$$
where
$$
\eqalignno{
Q_\sst{W}^{(0)}&=\Bigl({Z-N\over 2}\Bigr)\xivieo+\sqrt{3}\Bigl({Z+N\over 2}
\Bigr)\xiviez&(16{\rm a})\cr
\Delta Q_\sst{W}^{(n,\ p)}&=\coeff{1}{2}\bigl[\sqrt{3}\xiviez+\xivieo\bigr]
\rbra{I_0}\sum_{k=1}^{A}\coeff{1}{2}[1+\tau_3(k)]h(x_k)\rket{I_0}&\cr
&\quad\quad+\coeff{1}{2}\bigl[\sqrt{3}\xiviez-\xivieo\bigr]
\rbra{I_0}\sum_{k=1}^{A}\coeff{1}{2}[1-\tau_3(k)]h(x_k)\rket{I_0}&(16{\rm
b})\cr
\Delta Q_\sst{W}^{(s)}&=-\xivz\Bigl({\rhostr\over 4\mns}\Bigr)
\rbra{I_0}\sum_{k=1}^A\nabla_k^2 h(x_k)\rket{I_0}&(16{\rm c})\cr
\Delta Q_\sst{W}^{(I)}&=\lambda\xivieo\Bigl[\rbra{I_0}\sum_{k=1}^A
h(x_k)\tau_3(
k)
\rket{I_1}+ (I_1\leftrightarrow I_0)\Bigr]+\cdots\ \ \ ,&(16{\rm d})\cr}
$$
with $h(x)=f(x)-1$, and with
$\rbra{I_0}{\hat{\cal O}}\rket{I_0}$ denoting reduced matrix
elements of a nuclear operator $\hat{\cal O}$
in a nuclear ground state having nominal isospin $I_0$.
The terms in Eq. (16a) are those usually considered in analyses of
$\qw$. The term $\Delta Q_\sst{W}^{(n,\ p)}$ carries a dependence
on the ground-state neutron
radius, $R_n$. The impact of uncertainties in $R_n$ on the use of $\qw$ for
high-precision electroweak tests has been discussed in Refs.~[36, 37].
Eqs. (16c) and (16d) give, respectively, the leading contributions to $\qw$
from
$\GES$ and from isospin impurities in the nuclear ground state. In arriving
at Eq. (16), we have kept terms in $f(x)$ only up through quadratic order and
employed $R=r_o A^{1/3}$, $r_o\approx 1$~fm, for the nuclear radius. We have
shown explicitly only the contribution to $\Delta Q_\sst{W}^{(I)}$ arising from
the mixing of a single state of isospin $I_1$ into the ground state of nominal
isospin $I_0$ with strength $\lambda$. Additional
contributions to $\qw$ arising from the single-nucleon EM
charge radii are discussed elsewhere.$^{37}$

        According to Ref. [1], neglect of all but Eq. (16a) leads to the
prediction $\qwcs=-73.20-0.8\ S -0.005\ T$, so that a
0.7\% determination of $\qwcs$ would constrain $S$ to $|\delta S|\leq 0.6$.
As noted in Ref. [36], a 10\% uncertainty in $R_n$ would generate a 0.7\%
error in $\qwcs$. While hadron-nucleus scattering typically permits
a 5 - 10\% determination of $R_n$ for heavy nuclei,$^{19,\ 36}$ no experimental
information on $R_n$ for cesium isotopes presently exists. A series of PC and
PV electron scattering experiments on $^{133}$Cs could determine its neutron
radius to roughly 1\% accuracy.$^{6}$ In the meantime, one must rely on
nuclear model calculations of $R_n$. The scale of the associated theoretical
uncertainty in $\qwcs$ is presently the subject of debate.$^{37}$

{}From Eq. (16c), we find that an uncertainty in the strangeness radius
induces an error in the weak charge of $\delta\qwcs=-0.025 \delta\rhostr$.
For $\delta\rhostr$ on the order of the average value of Ref. [22], the
corresponding uncertainty in $\qwcs$ is slightly less than 0.1\%~, more than an
order of magnitude below the dominant theoretical error associated
with atomic structure$^{1,\ 4}$ and well below the level needed for
an interesting $\qwcs$ Standard Model test. As expected, the situation
differs sharply from that of PV electron scattering. Indeed,
a measurement of $\alrc$ would have to be
carried out at $|\ {\vec q}\ |\approx 30\ {\rm MeV}/c$ ---
roughly an order of magnitude
smaller than in the experiment of Ref. [3] -- to be equally insensitive to
$\GES$.

        We close with observations on the possibility of observing $\GES$
using PV experiments on muonic atoms. It has been noted recently that 1 - 10\%
measurements of PV observables for muonic boron may be feasible in the future
at PSI.$^{13,\ 38}$ Since the ratio of Bohr radii $a^e_0/a^\mu_0 = m_e/m_\mu
\sim 207$, the muon in these atoms is more tightly bound
for a given set of radial and angular momentum quantum numbers.
One might expect, then, an enhanced sensitivity to short-range
contributions to $\qw$, such as those associated with $R_n$ or $\rhostr$.
To analyze the latter possibility, we solve the Dirac equation for a muon
orbiting a spherically-symmetric nuclear charge distribution, keeping terms
involving $m_\mu$.$^{39}$ The result of this procedure is to make the
replacement $x_o=R/Z\alpha\to [3R/4m_\mu Z\alpha]^{1/2}$ in the function $h(x)$
in Eq. (16). The scale of $\Delta Q_\sst{W}^{(s)}$ is correspondingly
enhanced by $4m_\mu R/3Z\alpha\sim 4m_\mu r_oA^{1/3}/3Z\alpha$ over its
magnitude for an electronic atom. In the case of $^{133}$Cs, this
enhancement factor is $\approx 8$, making $Q_\sst{W}(\mu{\rm Cs})$
roughly as sensitive to $\rhostr$
as is $\alrep$.
The sensitivity of $\Delta Q_\sst{W}^{(s)}$
for a muonic lead atom is roughly two times greater than
$\Delta Q_\sst{W}^{(s)}(\mu{\rm Cs})$. For light
muonic atoms, on the other hand, the $\rhostr$ contribution is still
suppressed.
In the case of muonic boron, for example, uncertainties associated with
$\rhostr$ would not enter the parameters
$\xivp$ and $\xivn$ at an observable level.
Consequently, one must go to heavy muonic atoms. While the sensitivity of
the latter to $R_n$-uncertainties is also enhanced, these uncertainties could
be reduced through a combination of PC and PV elastic electron scattering
experiments.$^{6,\ 19}$
Given the simplicity of atomic structure calculations for muonic Cs or Pb
(essentially a one-lepton problem), the theoretical atomic structure
uncertainties entering $\qw$-determinations should not enter at a level
problematic for $\GES$ determinations. Thus, an experiment of this type could
complement PV electron scattering as a probe of strange quarks in the nucleon.
The remaining obstacle is the experimental one of achieving sufficient
precision. To this end, it would be desirable to find a heavy muonium
transition for which the PV signal is enhanced by accidental near degeneracies
between opposite-parity atomic levels.

\noindent{\bf 6. Conclusions}

        With any attempt at a precision electroweak test involving a low-energy
hadronic system, one must ensure that all sources of theoretical hadronic
physics uncertainties fall below the requisite level. The situation contrasts
with purely leptonic or high-energy electroweak tests. In the former case,
given a model of electroweak interactions, one can make precise and unambiguous
predictions for different observables, up to uncertainties associated with
unknown parameters ({\it e.g.}, $\mt$ and $\mh$) and with hadronic loops.
In the latter instance, strong-interaction uncertainties are controllable
through the use of a perturbative expansion and QCD. In the non-perturbative
low-energy regime, however, one must rely on the use of symmetries as well as
model estimates of, or independent experimental constraints on,
hadronic effects. The scale of uncertainty in a low-energy semi-leptonic
electroweak test, then, is set by experimental input and, where such is
lacking, any reasonable model estimate. In the foregoing discussion, we have
noted that completion of one or more PV electron scattering experiments
has the potential to complement atomic PV as a low-energy probe of new physics.
At present, however, experimental limits on nuclear dispersion corrections, as
well as theoretical predictions for the nucleon's strangeness form factors,
indicate that these two sources of hadronic physics uncertainty are too
large to make interesting electroweak tests possible with low-energy polarized
electrons. We have shown how a series of PV elastic scattering experiments
with $\nuc{4}{He}$ could reduce the uncertainty associated with the strangeness
radius below a problematic level. Achieving a better understanding of nuclear
dispersion corrections remains a challenge for both experiment and theory.
\goodbreak
\centerline{\bf ACKNOWLEDGEMENTS}

It is a pleasure to thank E.J. Beise, S.B. Kowalski, S.J. Pollock, and L.
Wilets
for useful discussions.
\bigskip
\vfill
\eject

\centerline{\bf REFERENCES}
\medskip
\item{1.}W.J. Marciano and J.L. Rosner,
{\it Phys. Rev. Lett. \bf 65} (1990) 2963.
\medskip
\item{2.}M.E. Peskin and T.
Takeuchi, {\it Phys. Rev. Lett. \bf 65} (1990) 964.
\medskip
\item{3.}P.A. Souder {\it et al.}, {\it Phys. Rev.
Lett. \bf 65} (1990) 694.
\medskip
\item{4.}S.A. Blundell, W.R. Johnson, and J. Sapirstein, \PRL{65} (1990)
        1411.
\medskip
\item{5.}M.C. Noecker {\it et al.}, {\it Phys. Rev. Lett. \bf 61} (1988) 310.
\medskip
\item{6.}M.J. Musolf and T.W. Donnelly, {\it Nucl. Phys. \bf A546} (1992)
        509.
\medskip
\item{7.}T.W. Donnelly, M.J. Musolf, W.M. Alberico, M.B. Barbaro, A. De Pace
and A. Molinari, {\it Nucl. Phys. \bf A541} (1992) 525; E. Hadjimichael,
G.I. Poulis, and T.W. Donnelly, {\it Phys. Rev. \bf C45} (1992) 2666.
\medskip
\item{8.}M.J. Musolf, T.W. Donnelly, J. Dubach, S.J. Pollock, S. Kowalski, and
E.J. Beise, MIT Preprint CTP \#2146 (1992), to be published.
\medskip
\item{9.}S.J. Pollock, {\it Phys. Rev. \bf D42} (1990) 3010.
\medskip
\item{10.}M.J. Musolf and B.R. Holstein, {\it Phys. Lett. \bf 242B} (1990) 461.
\medskip
\item{11.}W.J. Marciano and A. Sirlin, \PRD{29} (1984) 75.
\medskip
\item{12.} D. B. Kaplan and A. Manohar, \NPB{310} (1988) 527.
\medskip
\item{13.} P. Langacker, {\it Phys. Lett. \bf B256} (1991) 277.
\medskip
\item{14.} M.J. Musolf, MIT Preprint CTP \#2120 (1992), to appear in
proceedings of the 1992 CEBAF Summer Workshop, Newport News, VA.
\medskip
\item{15.} G. Feinberg, \PRD{12} (1975) 3575.
\medskip
\item{16.} J. D. Walecka, \NPA{285} (1977) 349.
\medskip
\item{17.} D. H. Beck, \PRD{39} (1989) 3248.
\medskip
\item{18.}R.G. Sachs, {\it Phys. Rev. \bf 126} (1962) 2256.
\medskip
\item{19.}T. W. Donnelly, J. Dubach and I. Sick, {\it Nucl. Phys.\/}
{\bf A503} (1989) 589.
\medskip
\item{20.}Spinor normalization convention is that of
Ta-Pei Cheng and Ling-Fong Li, {\it Gauge Theory of Elementary
Particle Physics}, Clarendon Press, Oxford, 1984.
\medskip
\item{21.}S. Galster {\it et al.}, {\it Nucl. Phys. \bf B32} (1971) 221.
\medskip
\item{22.}R. L. Jaffe, {\it Phys. Lett. \bf B229} (1989) 275.
\medskip
\item{23.}N. W. Park, J. Schechter and H. Weigel, {\it Phys. Rev.\/}
{\bf D43}, 869 (1991).
\medskip
\item{24.}W. Koepf, E.M. Henley, and S.J. Pollock, Inst. for Nucl. Theory
Preprint \#40427-15-N92 (1992).
\medskip
\item{25.}M. Burkardt and M.J. Musolf, to be published.
\medskip
\item{26.} These conditions differ somewhat from those expected to
be attainable at CEBAF. The Hall A spectrometers would detect electrons
at $\theta=12.5^\circ$ with a solid angle of $\Delta\Omega=0.016$ sr.
In an estimate of the statistical precision,
the smaller solid angle is compensated by an increase in the figure of
merit in the more forward direction. All other conditions being equal,
the constraints on $\GES$ attainable with two prospective Hall A measurements
would be similar to those shown in Fig. 4.
The projected beam polarization is likely to
reach no more than 80\% .
\medskip
\item{27.}S. D. Drell and S. Fubini, {\it Phys. Rev. \bf 113} (1959) 741.
\medskip
\item{28.}G. K. Greenhut, {\it Phys. Rev. \bf 184} (1969) 1860.
\medskip
\item{29.}N. Kalantar-Nayestanaki \etal , \PRL{63} (1989) 2032.
\medskip
\item{30.}J. L. Friar and M. Rosen, \AoP{87} (1974) 289.
\medskip
\item{31.}T. W. Donnelly, J. Dubach, and I. Sick, \PRC{37} (1988) 2320.
\medskip
\item{32.}J. Napolitano, \PRC{43} (1991) 1473.
\medskip
\item{33.}CEBAF proposal \# PR-91-010, J. M. Finn and P. A. Souder,
spokespersons.
\medskip
\item{34.}CEBAF proposal \# PR-91-017, D. H. Beck, spokesperson.
\medskip
\item{35.}CEBAF proposal \# PR-91-004, E. J. Beise, spokesperson.
\medskip
\item{36.}E.N. Fortson, Y. Pang, and L. Wilets, \PRL{65} (1990) 2857.
\medskip
\item{37.}S.J. Pollock, E.N. Fortson, and L. Wilets, Inst. for Nucl. Theory
Preprint \#40561-050-INT92-00-14 (1992).
\medskip
\item{38.} J. Missimer and L.M. Simons, PSI Preprint \#PR-90-03;
and {\it Phys. Rep. \bf 118} (1985) 179.
\medskip
\item{39.}We are indebted to Prof. L. Wilets for discussions on this
procedure.
\bigskip
\vfill
\eject
\centerline{\bf FIGURE CAPTIONS}
\medskip
\item{\rm Fig. 1.}Present and prospective constraints on $S,T$ parameterization
of non-standard physics from low- and intermediate-energy PV observables.
Short-dashed lines give present constraints from cesium atomic PV.$^{1,\ 4,\
5}$ Solid lines give constraints from a 1\% $\alrc$ measurement.
Long-dashed lines correspond to a 10\% determination of $\xivp$ from a
forward-angle measurement of $\alr(\evec p)$. For simplicity, it is assumed
that all experiments agree on  common central values for $S$ and $T$, so that
only the deviations from these values are plotted.
\medskip
\item{\rm Fig. 2.}Electron bremsstrahlung for electromagnetic (Fig. 2a,b) and
weak neutral current (Fig. 2c,d) scattering from a hadronic target. Target
bremsstrahlung is assumed to be negligible for low-energy (small recoil)
processes.
\medskip
\item{\rm Fig. 3.} Dispersion corrections to tree-level EM and NC
electron-nucleus scattering amplitudes. Here, $V,V'$ are any one of
the $Z^0, W^\pm, \gamma$ vector bosons and $\ket{i}$ ($\ket{f}$) are
initial (final) nuclear states.
\medskip
\item{\rm Fig. 4.} Constraints imposed on $\GES$ from prospective
PV elastic scattering experiments. Dashed-dot curves and solid curves give,
respectively, constraints from possible low- and moderate-$|Q^2|$ measurements
of $\alr(^{4}{\rm He})$. Dashed lines give constraints from series of
forward- and backward-angle $\alr(\evec p)$ measurements.
Panels (a) and (b) correspond to two models for $\GES$ discussed in the
text, where the canonical values of $(|\rhostr|, \lams)$ are indicated by
the large dot.
\par
\vfill
\end